\documentstyle[12pt]{article}
\def\beq{\begin{equation}}
\def\eeq{\end{equation}}
\def\bea{\begin{eqnarray}}
\def\eea{\end{eqnarray}}
\def\bq{\begin{quote}}
\def\eq{\end{quote}}

\def\als{\alpha_s}

\def\nnb{\nonumber}
\def\ga{\left(}
\def\dr{\right)}
\def\aga{\left\{}
\def\adr{\right\}}
\def\lb{\lbrack}
\def\rb{\rbrack}

\def\rar{\rightarrow}
\def\nnb{\nonumber}
\def\la{\langle}
\def\ra{\rangle}
\def\nin{\noindent}
\def\ba{\begin{array}}
\def\ea{\end{array}}

\begin{document}
\topmargin -1.5cm
\oddsidemargin -0.5cm
\evensidemargin -1.0cm
\pagestyle{empty}
\begin{flushright}
PM 95/07
\end{flushright}
\vspace*{1cm}
\begin{center}
\section*{QCD tests from $e^+e^- \rar I=1~$ hadrons data and\\
implication on the value of $\alpha_s$ from $\tau$-decays }
\vspace*{1.0cm}
{\bf S. Narison} \\
\vspace{0.3cm}
Laboratoire de Physique Math\'ematique\\
Universit\'e de Montpellier II\\
Place Eug\`ene Bataillon\\
34095 - Montpellier Cedex 05, France\\
\vspace*{1.0cm}
{\bf Abstract} \\ \end{center}
\vspace*{2mm}
\noindent
We re-examine the estimate of $\als$ and of the
QCD condensates from $e^+e^-\rar I=1$ hadrons data.
We conclude that $e^+e^-$ at low energies gives a value of $\Lambda$
compatible with the one from LEP and from tau inclusive decay.
Using a $\tau$-like inclusive process and QCD spectral sum rules,
we estimate the size of the $D$=4 to 9 condensates by a fitting
procedure {\it without invoking stability criteria}.
We find $\la \als G^2 \ra =(7.1\pm 0.7)10^{-2}$ GeV$^4$,
$\rho\als\la \bar uu\ra^2= (5.8\pm 0.9)10^{-4}$ GeV$^6$,
which confirm previous sum rules estimate based on stability
criteria. The corrections due to the $D=8$ condensates and to
instantons on the
vector component of $\tau$-decay are respectively
$\delta^{(8)}_1= -(1.5\pm 0.6)10^{-2}(1.78/M_\tau)^8$ and
$\delta^{(9)}_1=-(7.0\pm 26.5)10^{-4}(1.78/M_\tau)^9 $, which indicate
that the $\delta^{(8)}_1$ is
one order magnitude higher than the vacuum saturation value,
while the $D\geq 9$ instanton-like contribution to the
the vector component of the $\tau$-decay width is a negligible
correction.
We also show that, due to the correlation between the $D=4$ and
$1/M^2_\tau$ contributions in the ratio of the Laplace sum rules,
the present
value of the gluon condensate already excludes the recent estimate of
the $1/M^2_\tau$-term from FESR in the axial-vector channel.
Combining our non-perturbative results with
the resummed perturbative corrections to the $\tau$-width $R_\tau$,
we deduce from the present data
$\alpha_s(M_\tau)= 0.33\pm 0.03$.
\vspace*{3cm}
\begin{flushleft}
PM 95/07\\
April 1995
\end{flushleft}
\vfill\eject
\setcounter{page}{1}
 \pagestyle{plain}
\section{Introduction} \par
\nin
Measurements of the QCD scale $\Lambda$
and of the $q^2$-evolution of the QCD coupling are one of
the most important test of perturbative QCD. At present LEP
and $\tau$-decay data \cite{BETHKE}-\cite{SNA}
indicate that the value of $\alpha_s$ is
systematically higher than the one extracted from deep-inelastic
low-energy data.
The existing estimate of $\alpha_s$ from QCD
spectral sum rules \cite{SNB} \`a la SVZ \cite{SVZ} in $e^+e^-$ data
\cite{EID,BERT} also favours the low value of
$\alpha_s$ from deep-inelastic scattering
\footnote{However, new results of jet studies
in deep-inelastic $ep$-scattering at HERA for photon momentum transfer
$10 \leq Q^2$ [GeV$^2]\leq 4000$
give a value  of $\alpha_s$
\cite{H1} compatible with the LEP-average.}, which
is, however, in contradiction with the recent CVC-test
performed by \cite{PICH} using $e^+e^-$ data. It is therefore essential
to test
the reliability of the low-energy predictions before speculating on the
phenomenological consequences implied by the previous discrepancy.
\nin
 Deep-inelastic scattering
processes need a better control of the parton
distributions, the higher-twist and instanton-like
contributions in order to be competitive with LEP and tau-decay
measurements. In addition, perturbative corrections in these
processes should be pushed so far such that the remaining uncertainties
will only be due to the re-summation of the perturbative series at
large order. Indeed, the $\tau$-decay rate has been calculated including
the $\als^3$-term \cite{BNP},
 while an estimate \cite{KATA} and a measurement \cite{LEDI2}
of the $\als^4$ coefficient is done. Moreover, a resummation of the
$(\beta_1\alpha_s)^n$ of the perturbative series is now available
\cite{BENEKE}.
\nin
 The QCD spectral sum rule (QSSR) \cite{SNB} \`a la SVZ \cite{SVZ}
applied to
the $I=1$ part of the $e^+e^-\rar$ hadrons total cross-section has a
QCD expression very similar to the $\tau$-decay inclusive width, such
that on a theoretical basis, one can have a good control of it.
\nin
In a previous paper \cite{SN1}, we have derived in a model-independent
way the running mass of the strange quark from the difference between
the $I=1$ and $I=0$ parts of the $e^+e^- \rar$ hadrons total
cross-section. In this paper, we pursue this analysis by re-examining
the estimate of $\als$ and of the condensates including the
instanton-like and the {\it marginal} $D$=2-like operators obtained
from the $I=1$ channel of the $e^+e^-$ data.
In so doing, we re-examine the exponential Laplace
sum rule used by \cite{EID} in $e^+e^-$, which is a generalization of
the $\rho$-meson sum rule studied originally by SVZ \cite{SVZ}.
We also expect
that the Laplace sum rule gives a more reliable result than the FESR
due to the presence of the exponential weight factor which suppresses
the effects of higher meson masses in the sum rule. This is important
in the particular channel studied here as the data are very inaccurate
above 1.4--1.8 GeV, where the optimal result from FESR satisfies the
so-called heat evolution test  \cite{BERT,SN2,DOMI}.
That makes the FESR prediction strongly dependent on
the way the data in this region are parametrized, a feature which we
have examined \cite{PICH,SNACH} for criticizing the work
of \cite{TRUONG}. We also test the existing
and controversial results \cite{SN2,DOMI} of
the $D=2$-type operator obtained from QSSR. Combining our different
non-perturbative results with the recent resummed perturbative series
\cite{BENEKE}, we re-estimate the value of $\alpha_s$ from $\tau$-decays.
\section{$\als$ from $e^+e^-\rar I=1$ hadrons
data}
Existing estimates of $\als$ or $\Lambda$ from different aspects of
QSSR sum rules for $e^+e^- \rar I=1$ hadrons data
\cite{EID,BERT} lead to values much smaller than the present LEP and
$\tau$-decay measurements \cite{BNP}-\cite{SNA}.
However, such results contradict the
stability-test on the extraction of $\als$ from $\tau$-like
inclusive decay \cite{PICH} obtained using CVC in $e^+e^-$ \cite{GIL}
for different values of the $\tau$-mass.
In the following, we shall re-examine the
reliability of these sum rule results.
\nin
We shall not reconsider the result from FESR
\cite{BERT}
due to the drawbacks of this method mentioned previously, and also,
because the
FESR-analysis has been re-used recently \cite{SN2,DOMI} for a
determination of the $D=2$-type operator, which we shall come back
later on.
\nin
$\Lambda_3$ and the condensates have been
extracted in \cite{EID} from the Laplace sum rule:
\beq
{\cal L}_1\equiv \frac{2}{3}\tau\int_{4m^2_\pi}^{\infty}ds~e^{-s\tau}
R^{I=1}(s)
\eeq
and from its $\tau\equiv 1/M^2$ derivative:
\beq
{\cal L}_2\equiv \frac{2}{3}\tau^2\int_{4m^2_\pi}^{\infty}ds~s~e^{-s\tau}
R^{I=1}(s),
\eeq
where:
\beq
R^{I}\equiv\frac{\sigma(e^+e^-\rar I
\mbox{ hadrons})}{\sigma(e^+e^-\rar \mu^+\mu^-)}.
\eeq
In the chiral limit $m_u=m_d=0$, the QCD expressions
of the sum rule can be written as:
\beq
{\cal L}_i=1+\sum_{D=0,2,4,...}{\Delta_i^{(D)}}.
\eeq
The perturbative corrections can be deduced from the ones of $R^{I=1}$
obtained to order $\alpha_s^3$:
\beq
R^{I=1}(s)=\frac{3}{2}\aga 1+a_s+F_3a_s^2+F_4a_s^3+{\cal{O}}(a^4_s)\adr,
\eeq
where, for 3 flavours: $F_3=1.623$ \cite{KATAE}, $F_4=6.370$ \cite{GOR};
the expression of the running coupling to three-loop accuracy is:
\bea
a_s(\nu)&=&a_s^{(0)}\Bigg\{ 1-a_s^{(0)}\frac{\beta_2}{\beta_1}\log
\log{\frac{\nu^2}{\Lambda^2}}\nnb \\
&+&\ga a_s^{(0)}\dr^2\lb\frac{\beta_2^2}{\beta_1^2}\log^2\
\log{\frac{\nu^2}{\Lambda^2}}-\frac{\beta_2^2}{\beta_1^2}\log
\log{\frac{\nu^2}{\Lambda^2}}-\frac{\beta_2^2}{\beta_1^2}
+\frac{\beta_3}{\beta_1}\rb+{\cal{O}}(a_s^3)\Bigg\},
\eea
with:
\beq
a_s^{(0)}\equiv \frac{1}{-\beta_1\log\ga\nu/\Lambda\dr}
\eeq
and
$\beta_i$ are the  ${\cal{O}}(a_s^i)$ coefficients of the
$\beta$-function
in the $\overline{MS}$-scheme for $n_f$ flavours:
\bea
\beta_1&=&-\frac{11}{2}+\frac{1}{3}n_f \nnb \\
\beta_2&=&-\frac{51}{4}+\frac{19}{12}n_f\nnb \\
\beta_3&=&\frac{1}{64}\Big{[}-2857+\frac{5033}{9}n_f-\frac{325}{27}n^2_f
\Big{]}.
\eea
For three flavours, we have:
\beq
\beta_1=-9/2,~~~~\beta_2=-8,~~~~\beta_3=-20.1198.
\eeq
\nin
In the chiral limit, the $D=2$-contribution vanishes. It has also been
proved recently \cite{BENEKE} that renormalon-type
contributions induced by the resummation of the QCD series at large order
cannot induce such a term.
\nin
In the chiral limit, the $D=4$ non-perturbative corrections read
\cite{SVZ,BNP}:
\bea
\Delta^{(4)}_1 &=&\frac{\pi}{3}\tau^2\la \als G^2\ra \ga 1-\frac{11}{18}
\frac{\als}{\pi}\dr \nnb \\
\Delta^{(4)}_2 &=&-\Delta^{(4)}_1.
\eea
The $D=6$ non-perturbative corrections read \cite{SVZ}:
\bea
\Delta^{(6)}_1&=&-\frac{448\pi^3}{81}\tau^3\rho\la \bar uu\ra ^2\nnb \\
\Delta^{(6)}_2&=&-2\Delta^{(6)}_1.
\eea
We shall use the conservative values of the condensates \cite{SNB,BNP}:
\beq
\la \als G^2\ra =(0.06\pm 0.03)~{\mbox GeV^4}~~~~~~~
\rho\la \bar uu\ra ^2=(3.8\pm 2.0)10^{-4}~\mbox{GeV}^6,
\eeq
and $high$ values of $\Lambda$ from LEP and tau-decay data
\cite{BETHKE}-\cite{LEDI} for 3 flavours:
\beq
\Lambda_3= 375^{+105}_{-85} ~\mbox{MeV},
\eeq
corresponding to $\alpha_s(M_Z)=0.118\pm 0.06$.
\nin
The phenomenological side of the sum rule has been parametrized using
analogous data as \cite{EID} and updated using the data used in
\cite{PICH}. The confrontation of the QCD and the phenomenological sides
of the sum rules is done in Fig.1a
and in Fig. 2a for a giving value of $\Lambda_3=375$
MeV and varying the condensates in the range given previously. One can
conclude that one has a good agreement between the two sides of
${\cal{L}}_1$ for $M\geq 0.8$ GeV and of ${\cal{L}}_2$ for $M\geq
1.0\sim 1.2$ GeV. The effects of the condensates are important below
1 GeV for ${\cal{L}}_1$ and below 1.3 GeV for ${\cal{L}}_2$.
In Fig. 1b and Fig. 2b,
we fix the condensates at their central values and we vary $\Lambda_3$ in
the range given above. One can notice that a value of $\Lambda_3$ as high
as 525 MeV is still allowed by the data, while the shape of the QCD
curve for ${\cal{L}}_2$ changes drastically for a high value
of $\Lambda_3$. This phenomena is
not informative as, below 1 GeV, higher dimension condensates can
already show up and may break the Operator Product Expansion (OPE).
\nin
By comparing these results with the ones of \cite{EID}, one can notice
that our QCD prediction for ${\cal{L}}_1$ corresponding to the previous
set of parameters is as good as the fit of
\cite{EID}, while for that of ${\cal{L}}_2$, the agreement between the
two sides of the sum rule is obtained here at a slightly larger value of
M for high-values of $\Lambda_3$.
\nin
However, what is clear from our analysis is that the exponential Laplace
sum rules {\it do not exclude} values of $\Lambda_3$ obtained from
LEP and $\tau$-decay data, though they
cannot give a $more~precise$ information on the real value of $\Lambda_3$
if the condensates are left as free-parameters in the analysis.
It is also informative and reassuring, that our analysis supports
the value of $\Lambda_3$ obtained from $\tau$-decay and used via CVC
\cite{GIL} for $e^+e^-$, in order to
test the stability of the prediction for different values of the
$\tau$-mass \cite{PICH} from the expression which we shall
discuss below.
\section{The condensates from $\tau$-like decays}
\nin
In so doing, we shall work with the
vector component of the $\tau$ decay-like quantity deduced
from CVC \cite{GIL}:
\beq
R_{\tau,1}\equiv \frac{3\cos^2{\theta_c}}{2\pi\alpha^2}S_{EW}
\int_{0}^{M^2_\tau}ds~\ga 1-\frac{s}{M^2_\tau}\dr^2\ga 1+\frac{2s}
{M^2_\tau}\dr
\frac{s}{M^2_\tau}~\sigma_{e^+e^-\rar~I=1},
\eeq
where $S_{EW}=1.0194$ is the electroweak correction
from the summation of the leading-log contributions \cite{MARC}.
\begin{table*}[h]
\setlength{\tabcolsep}{1.5pc}
\caption{Phenomenological estimate of $R_{\tau,1}$  }
\begin{center}
\begin{tabular}[h]{c|c  }
\hline
& \\
$M_{\tau}$[GeV]& $R_{\tau,1}$  \\
& \\
\hline
& \\
$1.0$  & $1.608\pm 0.064$ \\
$1.2$ & $1.900 \pm 0.075$ \\
$1.4$ & $1.853 \pm 0.072$ \\
$1.6$ & $1.793 \pm 0.070$  \\
$1.8$ & $1.790 \pm 0.081$  \\
$2.0$ & $1.818 \pm 0.097$ \\
& \\
\hline
\end{tabular}
\end{center}
\end{table*}
\nin
 This
quantity has been used in \cite{PICH} in order to test the stability of
the $\alpha_s$-prediction obtained at the $\tau$-mass of 1.78 GeV. It
has also been used to test CVC for different exclusive channels
\cite{PICH,EID2}. Here,
we shall again exploit this quantity in order to deduce
{\it model-independent} informations on the
values of the QCD condensates. The QCD expression of $R_{\tau,1}$ reads:
\beq
R_{\tau,1}=\frac{3}{2}\cos^2{\theta_c}S_{eW}\ga 1+\delta_{EW}+
\delta^{(0)}+\sum_{D=2,4,...}{\delta^{(D)}_1}\dr,
\eeq
where
$\delta_{EW}=0.0010$ is the electroweak correction coming from the
constant term \cite{LI};
the perturbative corrections read \cite{BNP}:
\beq
\delta^{(0)}= \ga a_s\equiv \frac{\alpha_s(M_\tau)}{\pi}\dr+5.2023
a_s^2+26.366a_s^3+...,
\eeq
The $a_s^4$ coefficient has also been estimated to be about 103
\cite{KATA,LEDI2}, though we shall use
$(78\pm 25)a_s^4$ where the error reflects the uncalculated higher
order terms of the
$D$-function, while the first term is induced by the lower order
coefficients after the use of the Cauchy integration.
\nin
In the chiral limit $m_i=0$, the quadratic mass-corrections contributing
to $\delta^{(2)}_{1}$ vanish. Moreover, it has been proved \cite{BENEKE}
that the summation of the perturbative series cannot induce such a term,
while the one induced eventually by the freezing mechanism is safely
negligible \cite{ALTA,SN2}. Therefore, we shall
neglect this term in the first step of our analysis. We shall test,
later on, the internal consistency of the approach if a such term is
included into the OPE.
\nin
In the chiral limit $m_i=0$, the $D=4$ contributions read \cite{BNP}:
\beq
\delta^{(4)}_{1}=\frac{11}{4}\pi a_s^2\frac{\la\alpha_s G^2\ra}
{M^4_\tau},
\eeq
which, due to the Cauchy integral and to the particular $s$-structure of
the inclusive rate, the gluon condensate
starts at ${\cal O}(a_s^2)$. This is a great advantage compared with the
ordinary sum rule discussed previously.
The $D=6$ contributions read \cite{BNP}:
\beq
\delta_{1}^{(6)}
\simeq 7\frac{256\pi^3}{27}\frac{\rho\alpha_s\la \bar{\psi_i}\psi_i\ra^2}
{M^6_\tau},
\eeq
\nin
The contribution of the $D=8$ operators in the chiral limit reads
\cite{BNP}:
\beq
\delta^{(8)}_{1}=-\frac{39\pi^2}{162}\frac{\la \alpha_s G^2\ra^2}
{M^8_\tau}.
\eeq
The phenomenological parametrization of $R_{\tau,1}$ has been done using
the same data input as in \cite{SN2,PICH}. We give in Table 1 its value
for different values of the tau mass.
Neglecting the $D=4$-contribution which is of the order $\als^2$,
we perform a two-parameter fit of the data
for each value of $\Lambda_3$
corresponding to the world average value of $\als(M_Z)
=0.118\pm 0.006$ \cite{BETHKE,PDG}
and by letting the $D=6$ and $D=8$ condensates as free-parameters.
We show
the results of the fitting procedure in Table 2 for
different values of $\Lambda_3$.
\begin{table*}[h]
\setlength{\tabcolsep}{1.5pc}
\caption{Estimates of $d_6$ and $d_8$ from
$R_{\tau,1}$ for different values of $\Lambda_3$}
\begin{center}
\begin{tabular}[h]{c|c c }
\hline
& &\\
 $\Lambda_3$ [MeV]&$d_6$ [GeV$^6$]&$-d_8$ [GeV$^8$] \\
&& \\
\hline
&& \\
480 & $-.07\pm 0.43$&$1.15\pm 0.40$ \\
375 & $0.27\pm 0.34$&$0.69\pm 0.31$ \\
290 & $0.58\pm 0.29$&$0.83\pm 0.27$ \\
& \\
\hline
\end{tabular}
\end{center}
\end{table*}
\nin
The errors take into account the effects
of the $\tau$-mass moved from 1.6 to 2.0 GeV,
which is a negligible effect, and the one due to the data.
One can notice that the estimate of the $D=8$ condensates is quite
accurate, while the one of the $D=6$ is not very conclusive
for $\Lambda_3 \leq 350$ MeV. Indeed, only above this value,
one sees that the $D=6$ contribution is clearly positive as expected
from the vacuum saturation estimate . This fact also explains the
anomalous low value of $-d_8$ around this transition region. Using the
average value of $\Lambda_3$ in Eq. (13), we can deduce the result:
\beq
d_8\equiv M^8_\tau\delta^{(8)}_1=-(0.85\pm 0.18)\mbox{GeV}^8 ~~~~~~~
d_6\equiv M^6_\tau\delta^{(6)}_1=(0.34\pm 0.20)\mbox{GeV}^6,
\eeq
which we shall improve again later on once we suceed to fix the
value of $d_6$.
\section{The condensates from the ratio of the
Laplace sum rules}
\nin
Let us now improve the estimate of the $D=6$ condensates. In so doing,
one can remark that,
though there are large discrepancies in the estimate of the absolute
values of the condensates from different
approaches, there is a consensus in the estimate of
the ratio of the $D=4$ over the $D=6$ condensates
\footnote{We have multiplied the original error given by \cite{BORDES}
by a factor 10. The constraint obtained in \cite{MENES} is not
very conclusive as it leads to $r_{46}\leq $ 110 GeV$^{-2}$ and does
not exclude $\leq 0$ value of the condensates.}:
\bea
 r_{46}[\mbox{GeV}^{-2}]\equiv \frac{\la \als G^2 \ra}{\rho \als\la
\bar uu \ra ^2} =&  94.80\pm 23&~\cite{TARR} \nnb \\
 & 96.20\pm 35&~\cite{BERT} \nnb \\
 & 114.6\pm 16&~\cite{BORDES} \nnb \\
&92.50\pm 50&~\cite{SOLA}.
\eea
from which we deduce the $average$:
\beq
r_{46}=(105.9 \pm 11.9)~\mbox{GeV}^{-2}.
\eeq
We use the previous informations on $d_8$ and $r_{46}$
for fitting the value of the $D=4$
condensates from the ratio of the Laplace sum rules:
\beq
{\cal{R}}(\tau)\equiv \tau^{-2}\frac{{\cal{L}}_2}{{\cal{L}}_1},
\eeq
used previously by \cite{TARR} for a simultaneous estimate of the $D=4$
and $D=6$ condensates. We recall that the advantage of this quantity is
its less sensitivity to the leading order perturbative corrections. The
phenomenological value of ${\cal{R}}(\tau)$ is given in Fig. 2.
Using a one-parameter fit, we deduce:
\beq
\la \als G^2 \ra=(6.1\pm 0.7)10^{-2}~ \mbox{GeV}^{4}.
\eeq
Then, we re-inject this value of the gluon condensate into the tau-like
width in Eq. (14), from which we re-deduce the value of the $D=8$
condensate.
After a re-iteration of this procedure, we deduce our $final$ results:
\beq
\la \als G^2 \ra=(7.1\pm 0.7)10^{-2}~ \mbox{GeV}^{4}~~~~~~~
d_8=-(1.5\pm 0.6)~\mbox{GeV}^{8}.
\eeq
Using the mean value of $r_{46}$, we also obtain:
\beq
\rho\als\la \bar uu\ra^2= (5.8\pm 0.9)10^{-4} ~\mbox{GeV}^6.
\eeq
We consider these results as an improvement and a confirmation
of the previous result in Eq. (12). It is also informative to
compare these results with the ALEPH and CLEO II measurements of these
condensates from the moments distributions of the $\tau$-decay width.
The most accurate measurement leads to \cite{EXP}:
\beq
\la \als G^2 \ra =(7.8\pm 3.1)10^{-2}~ \mbox{GeV}^{4},
\eeq
while the one of $d_6$ has the same absolute value as previously but
comes with the wrong sign.
 Our value of
$d_8$ is in good agreement with the one $d_8\simeq -0.95$ GeV$^8$
in \cite{PICH,PICHA} obtained from the same quantity, but it
is about one order of magnitude higher than the
vacuum saturation estimate proposed by
\cite{BAGAN} and about a factor 5 higher than the CLEO II
measurement. However, it is lower by a factor 2$\sim$3 than the
FESR result from the vector channel \cite{SOLA}
\footnote{In the normalization of \cite{SOLA},
our value of $d_8$ translates into $C_8\la O_8\ra
=(0.18 \pm 0.04)$ GeV$^8$.} . The discrepancy with the
vacuum saturation indicates that this approximation is very crude, while
the one with the FESR is not very surprising. Indeed, the FESR approach
done in the vector and axial-vector channels \cite{BERT,SOLA}
tends {\it always to overestimate} the values of the QCD condensates. The
discrepancy with the CLEO II measurement can be understood from the wrong
sign of the $D=6$ condensate obtained there and to its correlation with
the $D=8$ one.
\section{Instanton contribution}
\nin
Let us now extract the size of the instanton-like contribution by
assuming that it acts like a $D\geq 9$ operator. A good place for doing
it is $R_{\tau,1}$
as, in the Laplace sum rules, this contribution is suppressed by a 8!
factor. Using the previous values of the $D=6$ and $D=8$ condensates,
we deduce:
\beq
 \delta^{(9)}_1 =-(7.0\pm 26.5)10^{-4}(1.78/M_\tau)^9,
\eeq
which, though inaccurate indicates that
the instanton contribution is negligible
for the vector current and has been overestimated
in \cite{BRAUN} ($\delta_V^{inst}\approx 0.03\sim 0.05$).
Our result supports the negligible effects found from an
alternative phenomenological \cite{KART}
($\delta_V^{inst}\approx 3\times 10^{-3}$) and theoretical \cite{POR}
($\delta_V^{inst}\approx 2\times 10^{-5}$)
analysis. Further cancellations in the sum of the vector and axial-vector
components of the tau widths are however expected \cite{BRAUN,KART}
($\delta^{inst}\approx \frac{1}{20}\delta_V^{inst}$).
\section{Test of the size of the $1/M^2_\tau$-term}
\begin{table*}[h]
\setlength{\tabcolsep}{1.5pc}
\caption{Estimates of $\la \als G^2\ra$ from $R(\tau)$
for different values of $d_2$}
\begin{center}
\begin{tabular}[h]{c c | c c}
\hline
&& &\\
 $d_2$ [GeV]$^2$ \cite{SN2}&$\la \als G^2\ra 10^2$ [GeV$^4$]&
$-d_2$ [GeV]$^2$ \cite{DOMI}&$\la \als G^2\ra 10^2$ [GeV$^4$]\\
&&& \\
\hline
&&& \\
0.03 & $7.8\pm 0.5$&$$&$$ \\
0.05 & $8.1\pm 0.5$&$0.2$&$3.2\pm 0.29$ \\
0.07 & $8.6\pm 0.5$&$0.3$&$1.2\pm 0.29$ \\
0.09 & $9.1\pm 0.5$&$0.4$&$-0.7\pm 0.6$ \\
&&& \\
\hline
\end{tabular}
\end{center}
\end{table*}
\nin
Let us now study the size of the $1/M^2_\tau$-term.
{}From the QCD point of
view, its possible existence from the resummation of the PTS due to
renormalon contributions
\cite{ALTA} has not been confirmed \cite{BENEKE}, while
some other arguments \cite{ALTA,SHIF} advocating its existence are not
convincing.
Postulating its existence (whatever its origin!),
\cite{SN2} has estimated
the strength of this term by using FESR and the ratio of moments
${\cal{R}}(\tau)$. As already mentioned earlier,
the advantage in working with the ratio is that
the leading order perturbative corrections disappear such that in a
$compromise$ region where the high-dimension condensates are still
negligible, there is a $possibility$ to pick up the
$1/M^2_\tau$-contribution. Indeed,
using usual stability criteria and {\it allowing a large range of
values around
the optimal result}, \cite{SN2} has obtained the $conservative$ value:
\beq
d_2 \equiv C_2 \equiv \delta^{(2)}_1 M^2_\tau
\simeq {(0.03\sim 0.08)~\mbox{GeV}^2},
\eeq
while the estimate of \cite{SN2} from FESR applied to the vector current
has not been very conclusive, as it leads to the inaccurate value:
\beq
d_2
\simeq {(0.02\pm 0.12)~\mbox{GeV}^2}.
\eeq
\nin
However, the recent FESR analysis from the axial-vector current
obtained at about the same value of the continuum threshold $t_c$
satisfying the so-called evolution test \cite{BERT}, is
$surprisingly$ very precise \cite{DOMI}
and disagrees in sign and magnitude with
our previous estimate from the ratio of moments. Assuming
a quadratic dependence in $\Lambda_3$, the result of \cite{DOMI} reads:
\beq
d_2\simeq -{(0.3\pm 0.1)~\mbox{GeV}^2},
\eeq
which is $surprisingly$ very precise taking into account the fact that
the spectral function of
the axial-vector current is not better measured than that of the vector
current. We test the reliability of this result, by remarking that $d_2$
(if it exists!) is strongly correlated to $d_4$
in the analysis of the ratio of Laplace sum rules
$R(\tau)$, while it is not the case between $d_2$ or $d_4$
with $d_6$ and $d_8$. Using our previous values of $d_6$
and $d_8$, one can study the variation of $d_4$ given the
value of $d_2$. The results given in Table 3 indicate that the
present value of the gluon condensate excludes the value of $d_2$
in Eq. (31) and can only permit a negligible fluctuation
around zero of this contribution, which is should not exceed the value
$0.03\sim 0.05$. This result rules out the possibility to have
a sizeable $1/M^2_\tau$-term \cite{ALTA,SHIF} and justifies its
neglection in the analysis of the $\tau$-width. More precise measurement
of the gluon condensate or more statistics in the $\tau$-decay data
will improve this constraint.
\section{Sum of the non-perturbative corrections to $R_\tau$}
\nin
Using our previous estimates, it is also informative to
deduce the sum of the non-perturbative contributions to the decay widths
of the observed heavy lepton of mass 1.78 GeV. In so doing, we add the
contributions of operators of dimensions $D=4$ to $D=9$ and we neglect
the expected small $\delta^{(2)}$-contribution.
\nin
For the vector
component of the tau hadronic width, we obtain
\footnote{We have used, for $M_\tau=1.78$ GeV, the conservative values:
$\delta^{(9)}_V\approx -\delta^{(9)}_A\simeq -(0.7\pm 2.7)10^{-3}$
and $\delta^{(9)}\approx 1/20\delta^{(9)}_V$ \cite{BRAUN}.}:
\beq
\delta^{NP}_V\equiv \sum_{D=4}^{9}\delta^{(D)}_1=(2.38\pm 0.89)10^{-2},
\eeq
while using the expression of the corrections for the axial-vector
component given in \cite{BNP}, we deduce:
\beq
\delta^{NP}_A=-(7.95\pm 1.12)10^{-2},
\eeq
and then:
\beq
\delta^{NP}\equiv \frac{1}{2}(\delta^{NP}_V+\delta^{NP}_A)
=-(2.79\pm 0.62)10^{-2},
\eeq
Our result confirms the smallness of the non-perturbative corrections
measured by the ALEPH and CLEO II groups \cite{EXP}:
\beq
\delta^{NP}=(0.3\pm 0.5)10^{-2},
\eeq
though the exact size of the experimental number is not yet very
conclusive.
\section{Implication on the value of $\alpha_s$ from $R_\tau$}
\begin{table*}[t]
\setlength{\tabcolsep}{1.5pc}
\caption{QCD predictions for $R_\tau$ using the
$contour~ coupling$-expansion}
\begin{center}
\begin{tabular}[t]{c| c c c c}
\hline
&&&& \\
 $\alpha_s(M_{\tau})$&$a_s^3$ & $a_s^4$ &$a_s^6$&$a_s^8$\\
&&&& \\
\hline
&&&& \\
0.26 & $3.364\pm 0.022$&$3.370$&$3.380\pm 0.019$&3.381 \\
0.28 & $3.402\pm 0.024$&$3.411$&$3.426\pm 0.019$&3.426 \\
0.30 & $3.442\pm 0.026$&$3.453$&$3.474\pm 0.021$&3.472 \\
0.32 & $3.484\pm 0.030$&$3.498$&$3.526\pm 0.023$&3.520 \\
0.34 & $3.526\pm 0.033$&$3.546$&$3.582\pm 0.031$&3.568\\
0.36 & $3.571\pm 0.040$&$3.594$&$3.640\pm 0.045$&3.613\\
0.38 & $3.616\pm 0.040$&$3.645$&$3.706\pm 0.069$&3.655\\
0.40 & $3.664\pm 0.040$&$3.700$&$3.775\pm 0.108$&3.685\\
&&&& \\
\hline
\end{tabular}
\end{center}
\end{table*}
\nin
Before combining the previous non-perturbative results
 with the perturbative correction to $R_\tau$, let us test the accuracy
of the resummed $(\alpha_s\beta_1)^n$ perturbative result of
\cite{BENEKE}. In so doing, we
fix $\alpha_s(M_\tau)$ to be equal to 0.32 and we compare the resummed
value of $\delta^{(0)}$ including the $\alpha_s^3$-corrections with the
one
where the coefficients have been calculated in the $\overline{MS}$-scheme
\cite{KATAE}.
We consider the two cases where
$R_\tau$ is expanded in terms of the usual coupling $\als$
or in
terms of the {\it contour coupling} \cite{LEDI}.
In both cases, one can
notice that the approximation used in the resummation technique tends
to overestimate the
perturbative correction by about 10$\%$. Therefore, we shall reduce
sytematically by 10$\%$, the
prediction from this method from the $\alpha_s^5$ to
 $\alpha_s^9$ contributions. We shall use the coefficient 27.46 of
$\alpha_s^4$ estimated in \cite{KATA,LEDI2}. Noting that, to the
order where the perturbative series (PTS) is estimated, one has alternate
signs in the PTS, which is an indication for reaching the
asymptotic regime. Therefore, we can
consider, as the best estimate of the resummed PTS, its value at the
minimum. That is reached, either for truncating the PTS by including the
$\alpha_s^6$ or the $\alpha_s^8$ contributions. The corresponding
value of $R_\tau$ including our non-perturbative contributions in
Eq. (34) is given in Table 4. We show for comparison the value of
$R_\tau$ including the $\alpha_s^3$-term, where we have
used the perturbative estimate in \cite{PICHA}
(the small difference with the previous
papers \cite{LEDI,PICH,PICHA,SNA,SNACH}
comes from the different non-perturbative term used here),
while the error quoted there comes from the na\"\i ve estimate
$\pm 50 a_s^4$. However, one can see that the estimate
of this perturbative error has taken properly the inclusion of the higher
order terms, while the truncation of the series at $\alpha_s^3$
already gives a quite good evaluation of the PTS. One can also notice
that there is negligible difference between the PTS to order
 $\alpha_s^6$ and $\alpha_s^8$ for small values of
$\als$, while the difference increases for larger values.
We consider as a final perturbative estimate of
$R_\tau$ the one given by the PTS including the $\alpha_s^6$-term at
which we encounter the first minimum. The error
given in this column is the sum of the non-perturbative
one from Eq. (34) with the perturbative conservative
uncertainty, which we have estimated like the
effect due to the last term i.e $\pm 34.53(-\beta_1 a_s/2)^6$
at which the minimumm is reached, which is a legitime procedure
for asymptotic series \cite{HARDY}. We have also added to the latter
the one due to the small fluctuation of the
minimum of the PTS from the inclusion of
the $\alpha_s^6$ or $\alpha_s^8$-terms. One can notice that for $\als\leq
0.32$, the error in $R_\tau$ is dominated by the non-perturbative one,
while for larger value of $\als$, it is mainly due to the one from the
PTS. Using the value of $R_\tau$ in Table 4, we deduce:
\beq
\alpha_s(M_\tau)= 0.33\pm 0.030,
\eeq
where  we have used the experimental average \cite{PDG}:
\beq
 R_\tau= 3.56 \pm 0.03.
\eeq
Our result from the optimized resummed PTS is
in good agreement with the most recent estimate obtained to order
$\alpha_s^3$ \cite{PICHA,EXP,SNA}:
\beq
\alpha_s(M_\tau)= 0.33 \pm 0.030.
\eeq
\section{Conclusion}
\nin
Our analysis of the isovector component of the $e^+e^-\rar $ hadrons
data has shown that there is a consistent picture on the extraction of
$\als$ from high-energy LEP and low-energy $\tau$ and $e^+e^-$ data.
\nin
It has also been shown that the values of the condensates obtained from
QCD spectral sum rules based on {\it stability criteria} are reproduced
 and improved by fitting the $\tau$-like decay widths and the
ratio of the Laplace sum rules. Our estimates are in good agreement with
the determination of the condensates from the the $\tau$-hadronic
width moment-distributions \cite{EXP}, which needs to be improved from
accurate mesurements of the $e^+e^-$ data or/and for more data sample of
the $\tau$-decay widths which can be reached at the $\tau$-charm factory
machine.
\nin
Finally, our consistency test of the effect of the $1/M^2_\tau$-term,
whatever its origin, does not
support the recent estimate of this quantity from FESR axial-vector
channel \cite{DOMI} and only permits a small fluctuation around zero due
to its strong correlation with the $D=4$ condensate effects in the ratio
of Laplace sum rules analysis,
indicating that it cannot affect in a sensible way the accuracy of
the determination of $\als$ from tau decays.
\nin
As a by-product, we have reconsidered the estimate of $\als(M_\tau)$
from the $\tau$-widths taking into
account the recent resummed result of the perturbative series. Our result
in Eq. (36) is a further support of the existing estimates.
\section*{Acknowledgements}
It is a pleasure to thank A. Pich for exhanges and for carefully
reading the manuscript.
\vfill\eject
\section*{Figure captions}
\nin
{\bf Fig. 1a:} The Laplace sum rule ${\cal{L}}_1$ versus the sum rule
parameter M. The dashed curves correspond to the experimental data. The
full curves correspond to the QCD prediction for $\Lambda_3$=375 MeV,
$\la \als G^2 \ra=0.06 \pm 0.03$ GeV$^4$ and $\rho\la\bar uu\ra^2=
(3.8\pm 2.0)10^{-4}$ GeV$^6$.
\vspace*{0.5cm}
\nin
{\bf Fig. 1b:} The same as Fig. 1a but for different values of
$\Lambda_3$ and for
$\la \als G^2 \ra=0.06 $ GeV$^4$ and $\rho\la\bar uu\ra^2=
3.8~10^{-4}$ GeV$^6$.
\vspace*{0.5cm}
\nin
{\bf Fig. 2a:} The same as Fig. 1a but for  ${\cal{L}}_2$.
\vspace*{0.5cm}
\nin
{\bf Fig. 2b:} The same as Fig. 1b but for  ${\cal{L}}_2$.
\vspace*{0.5cm}
\nin
{\bf Fig. 3:} Experimental value of the ratio of Laplace sum rules
  ${\cal{R}}(\tau)$ versus the sum rule variable $\tau\equiv 1/M^2$.
\vfill \eject

\end{document}